# Collapses of underground cavities and soil-structure interactions: influences of the position of the structure relative to the cavity

M. Caudron, M. Al Heib
*INERIS, Direction des Risques du Sol et du sous-sol, Parc technologique ALATA, BP2. F-60550 Verneuil en Halatte, France.*

F. Emeriault
*INSA-Lyon, LGCIE, F-69621, FRANCE*

ABSTRACT: This paper is focused on subsidence of small extend and amplitude caused by tunnel boring or the collapse of underground cavities, whether natural or man-made. The impact of the ground movements on existing structures is generally dramatic. It is therefore necessary to accurately predict these movements (settlements and horizontal extension or compression displacements). Even though it is obvious that the overall stiffness and weight of the structure influences the size and shape of the soil movement, the main features of this soil-structure interaction phenomenon are not well established. Caudron et al. (2006) developed an original small-scale physical model to take the soil-structure interaction into account. It is based on the use of the frictionnal Schneebeli material (assembly of small diameter rods) and a modified version including cohesion in order to reproduce a cohesive layer above a cavity. The displacements of the soil are obtained from digital images processing by particle image velocimetry (PIV). Interesting results were obtained, probing that the soil-structure interactions could be analysed by this experimental model. This article is focused in a first part on the influence of the position of the structure with respect to the cavity position. Consequences on the areas mainly concerned by horizontal compression or extension of the structure are determined. It appears that the stresses induced in the building are a superposition of several elementary loading (sum of the effetcs of the slope, the horizontal deformations and the curvature). The second part concerns the effect of the relative soil/structure stiffness on the ground movement during a cavity collapse by considering a second model of structure with similar dimensions but more flexible.

## 1 Introduction

The subsidences are caused by the collapse of anthropic or natural underground cavities. The sudden nature of this phenomenon can be very prejudicial for the structures and infrastructures on the surface and for the population. Several research projects are focused on the study of the ground-structure interactions phenomena under the effect of the soil movements (Standing 2008, Lee et al. 2007, Sung et al. 2006, Abbass 2004, Deck 2002, Burd et al. 2000, Nakai et al. 1997).
While previous works (Caudron 2007) had allowed to shed some light on the interest of taking into account the soil-structure interaction, this paper will focus on the influence of some parameters on the structure behavior. The concerned parameters are the position of the structure relative to the cavity and the stiffness of the structure.
This study is achieved by the mean of a bidimensionnal small-scale physical model previously presented and discussed in Caudron et al. (2006). The main facts will be reminded in a first part. Then the aim of the parametric study will be defined and the different test configurations related to full scale observations. The third part is focused on the results from these tests and on the possible interpretation relatively to the behaviour of the soil and the structure.

## 2 The physical model

Caudron et al. (2006) have widely presented the small-scale physical model: design and limits of the model. The first step in the definition of the small-scale physical model was the laws of similarities. The assumption is made that the thermal effect has no influence on the series of tests, making the problem a bit simpler. Then three scale factors must be determined in order to fix the whole relation between the full-scale case study and the small-scale model. These scale factors concerned the gravity, the density and the length. The tests are performed under normal gravity, the corresponding scale factor is thus 1. The analogical soil has a unit weight of 65kN/m$^3$ (see below for more details), the scale factor on density is equal to 3. The last scale factor concerns the length. It has been fixed to 1/40 in order to have a test bed with convenient dimensions. From this point, all the other scale factors can then be deduced from the laws of similarities and these three values. Therefore, it is not possible to respect all the rules of similarities and more particularly those concerning the stress states. The results of the small scale tests will then be only qualitative and not quantitative.

The small-scale model uses the analogical soil of Schneebeli, composed of 60 mm long metallic rods with given diameters (3, 4 and 5 mm in equal weight-proportion). In order to allow the creation of a cavity, the main requirement of the previous research was to be able to represent a layer of coherent soil over the cavity. This was achieved by soaking the concerned rods in an aqueous solution of glue. Then these rods are manually put on the test bed in the desired place and dried until complete dehydration. The physical and mechanical characteristics of both materials are presented in table 1.

The test creation process is composed of different phases. The test bed is composed of two parts: the U-shaped frame (figure 1) and the apparatus used to create the cavity (figure 2). It allows an easy and repetitive creation of the cavity. Moreover, cavities of different sizes can be stepwise created: from 25 mm up to 250 mm in width with a maximum of 10 steps and from 25 mm to 100 mm in height. Ten 25 mm wide moving parts compose the cavity. Each one is equipped with a small force sensor on its top end. This will allow observing the variation of the applied forces developping within the above soil mass.

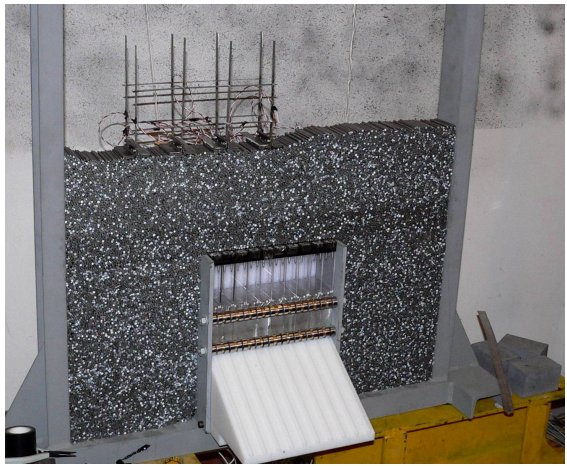

Figure 1. View of the small-scale experimental model (apparatus for the creation of the cavity at the centred bottom, Schneebeli material, building model on the ground surface).

Table 1. Mechanical characteristics of the analogical soils (prototype scale and full scale).

|  | Density | E (MPa) | φ (°) | c (kPa) |
|---|---|---|---|---|
| **Cohesionless soil** (prototype scale) | 6.5 | 4-8 | ~26 | 0 |
| Full scale | 2.2 | 50-100 | ~26 | 0 |
| **Coherent soil** | 6.5 | 4-8 | ~26 | ~16 |
| Full scale | 2.2 | 50-100 | ~26 | ~200 |

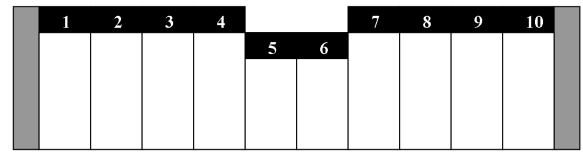

Figure 2. Experimental apparatus for the creation of the cavity with the ten different prisms (the moving parts 5 and 6 are lowered, and then number 4 and 7, and so on).

The Schneebeli rods are then put on the test bed in order to follow the chosen geometry and stratification with the coherent material. If needed the building model is placed on the ground surface.

This structure (figure 3) is composed of beams and columns. It is supposed to be representative of a small apartment building. The characteristics of the beams have been determined in order to get a bending stiffness respecting the laws of similitude and the corresponding scale factors. Thus it has been decided to respect as far as possible the EI and EA products (with E the Young modulus of the used material, A and I respectively the cross-section and the moment of inertia). The table 2 presents the values chosen for the different components of both building models used in the following tests.

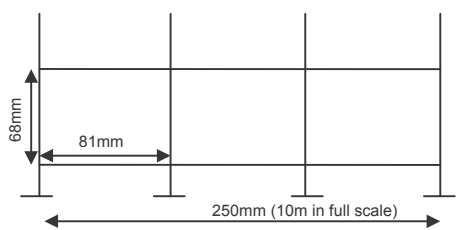

Figure 3. Schematic view of the building model.

Table 2. Stiffness values for both building models.

| Structure type | E A (N) | E $I_z$ (Nm²) |
|---|---|---|
| Steel structure | $1{,}34 \cdot 10^6$ | $175 \cdot 10^9$ |
| PVC structure | $66 \cdot 10^3$ | $88 \cdot 10^9$ |

The last step is the setup of the instrumentation. A PIV-DIC[1] procedure is used to follow the movements of the particles in the soil mass. White et al. (2003) clearly showed the pros and cons of such a method against photogrammetry. This solution allowed to get some good results in equivalent works (Jenck et al. 2007, Dolzhenko et al. 2001). The precision corresponding to the setup and equipment used is approximatively of 20 µm for a single couple of images. The building model is equipped with twelve strain gauges (see figure 5) and LVDT sensors (two on each foundation measuring thus the vertical displacement and the tilt).

## 3  The parametric study

Two parameters have been chosen for the study. They are both related to the structure, thus the setup of the ground does not vary for all the given tests (50 mm of coherent material then 250 mm of cohesionless analogical soil). This parametric study is thus only focused on the influence of the structure on the ground behaviour. The two chosen parameters are the location of the building model relative to the cavity and the stiffness of the building. There may be several other parameters that should have been studied. However these two parameters concern in a general way all the cavity related damages observed on buildings.

Depending on the size and depth of the cavity, parts of the subsidence trough would present some compressive horizontal strains while the other parts would be in extension. Moreover the size of the chosen building model is close to twice the value of *i*, the half distance between the points of inflexion which is correlated to the width of the trough. Therefore, to analyse the effect of the position of the structure, three different positions have been selected: a building model completely in the compression area (position n°1), then completely in the extension area (position n°2) and finally on both areas (position n°3) (figure 4). Due to the size of the building (around 240mm) and the extension of the subsidence trough (around 600mm), the third position is not independent from the two others. This position should present some disturbed results due to the relatively close dimensions of the building and of the studied phenomenon. For the two other positions, the building is completely in an homogeneous area regarding the horizontal deformation, the behaviour should easier to understand.

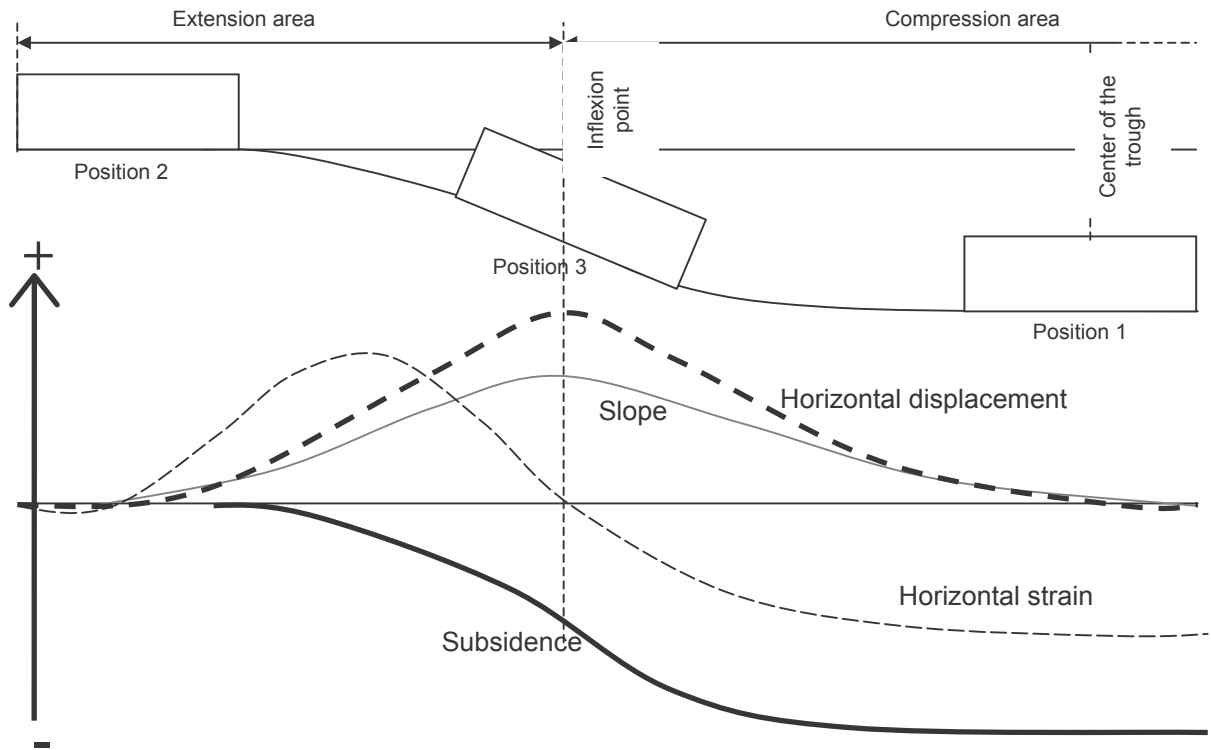

Figure 4. Schematic view of different positions used for the building model, relative to the shape of the trough and the corresponding soil movements.

For the second part of the parametric study, the stiffness of the building is inspired from two types of structures: masonry and framed structure. The first one implies a relatively stiff building while the second allows a more flexible behaviour. Two building models have been thus designed and prepared. The first one is trying to respect the relevant scale factor on the bending stiffness of a masonry building. The second one is associated to a bending stiffness two times smaller (table 2). This is achieved by the use of different materials and different pile sections for the two building models. Both building models are then studied with a test bed. It allows applying on a structure a determined set of boundary conditions. As illustrated by the figure 5, three foundations of the building model are fixed while a specific displacement can be imposed on the last one. Considering only the field of elastic strain, it is then possible to compare the behaviour of this model with a numerical model based on a linear elastic constitutive law. Results of this experiment and comparisons are plotted and analysed in Section 4.
The actual behaviour of both buildings has been determined and compared to the ideal expected behaviour.
In the parametric study, for each configuration, two identical tests are achieved to confirm the results.

---
[1] Particle Image Velocimetry – Direct Image Correlation

## 4 Results and interpretation

### 4.1 Results of the tests on both building models

We remind that the purpose of these tests is to achieve the characterisation of the behaviour of both building models under different sets of loadings.
Several tests are performed with both buildings (illustrated for example on figure 5for an imposed vertical displacement on the third foundation). Vertical or horizontal displacements are imposed on one of the foundations and the strains measured with the strain gauges. The behaviour of each building is elastic: no unrecoverable deformation can be observed during the tests (figure 6) but the observed measured deformations are quite different from the "theory".
A numerical model is used as representative of the theoretical linear elastic behaviour. The difference concerns mainly the values of the strains that should appear in the small-scale model. A factor varying from 0,5 to 5 is observed between the values resulting from the numerical simulation and those measured on the physical model. At this time, the only possible explanations for these discrepancies are the use of a linear elastic law for the material, the assumptions made on the geometry of the building for the modelling (constant cross-section and moment of inertia) and on the connections between beams and columns (perfect end restraint).

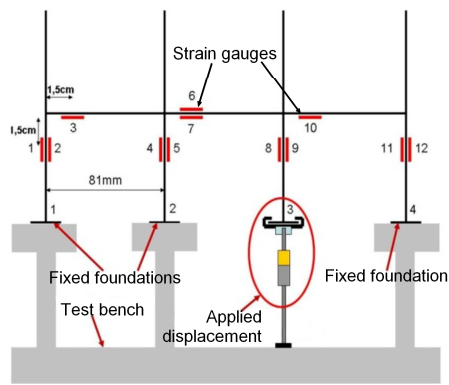
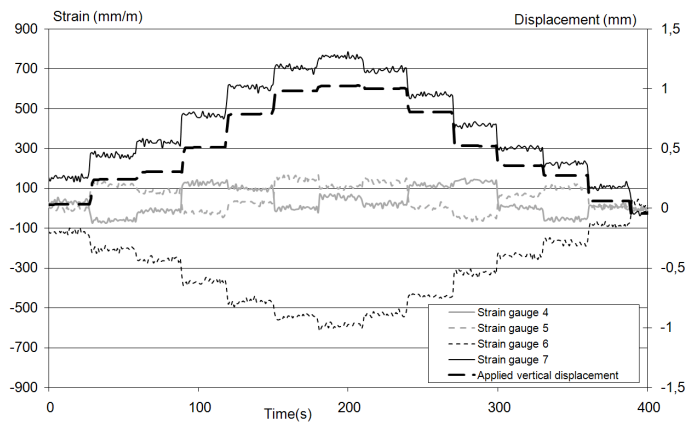

Figure 5. Schematic view of the setup used to test the behaviour of both building model.

Figure 6. Strain measured in the building model for different applied vertical displacement (loading and unloading).

### 4.2 Soil displacement in greenfield condition

A comprehensive series of tests have been performed in greenfield condition using the small-scale physical model (see Caudron et al. 2006, Caudron et al. 2007). The cavity collapse results in a subsidence trough whose characteristics are quite well known. The main parameters like the maximum settlement, the width of the trough can be predicted with a relatively good precision, knowing only the different geometrical parameters and the nature of the overburden.

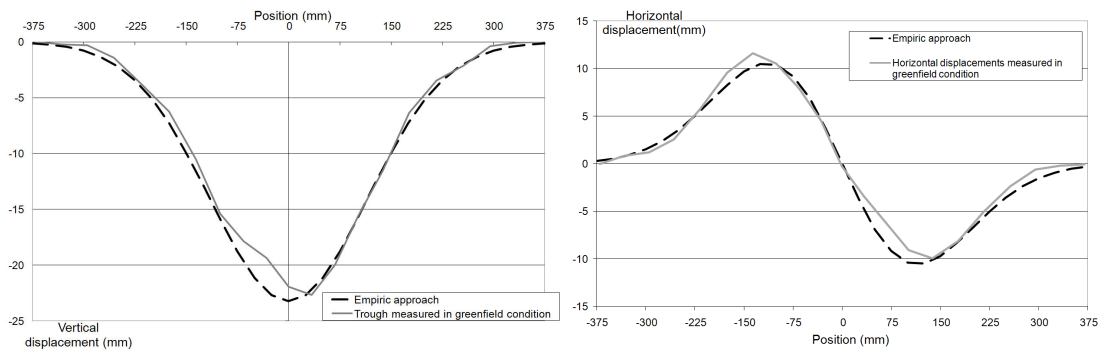

Figure 7. Comparison between the empirical approach and the observed behaviour on the physical model.

Caudron et al. (2007) have shown that the shape of the subsidence trough can be approximated by Peck's (1969) equation. The main hypothesis is made on the continuous nature of the trough. This approach is not able to take into account any discontinuities like those which may arise during a collapse. Figure 7 shows the agreement between experimental results and Peck's empirical approach. Despite some natural dispersion, the displacements provided by this approach can therefore be considered as the reference behaviour.

### 4.3 Tests relative to the position of the building

The three considered positions are shown in figure 4. For each, the results coming from two different tests are plotted in term of vertical and horizontal displacements (figures 8a to 8c). Several observations can then be made:
- The shape of the subsidence trough is very dependant on the position of the building (especially for case 2);
- The maximum settlement is smaller when the building is centred on the cavity;
- The slope of the trough seems to be more important in case of a centred building;
- While the horizontal displacements are quite different depending on the position of the building, we cannot observe any real variation on the horizontal deformations.

Considering the influence of the location of the building, it appears clearly that its nature depends greatly on the position of the structure: in the extension area or in the compression area.

In fact, when the building is centered on the cavity, its high stiffness comparatively to the stiffness of the ground tends to negate the horizontal displacements. It presents a rigid-body-like behaviour. The settlements are averaged within the location of this structure.

On the other hand, when the building is partially over the cavity, its influence of the soil displacements is different. While it slightly affects the shape of the subsidence trough, the main effect appears on the horizontal displacements, in a similar manner as observed by Caudron et al. (2007). They are reduced under the structure due to its important stiffness. They are undisturbed on the rest of the model test. This is confirmed by both tests realized in the same conditions which show similar results.

The second point of the results concerns the strains in the building model. The strain gauges placed on it allow us to observe the state variation during each test. Although the measured values present a certain coherence, we observe too much variations between the different tests for them to be usable. The behaviour of the building is quite different and thus doesn't allow to draw any real conclusion.

### 4.4 Tests relative to the building stiffness

As mentionned previously, a second building model was developed using a bending stiffness 0.5 times the stiffness of the initial building model. A set of two tests using the same initial conditions with the building in position n°3 is then achieved. The same measurement equipment is used as previously. Both tests present very similar results, so only one is presented on figure 9 in order to allow an easy comparison with the empirical approach corresponding to the Greenfield condition and the results of the test with the building of normal stiffness.

The results can be analysed separately: the soil movements first and then the building behaviour. The subsidence trough and the horizontal displacements of the ground surface (figure 9) present very few differences with previous results on the rigid structure. The shape is the same, as the settlement values if we make abstraction from the right part of the subsidence trough relative to the second test (this is a consequence of a bad failure caused by the apparatus used to create the cavity). For the horizontal displacement, the main difference is the absence of disturbance in the curve comparatively to the results obtained for the steel structure. It seems to be due to the use of a more flexible structure model which may admit higher strains that the previous one. Thus the computation of the slope and the horizontal strains is more precise. Comparatively to the tests with the stiffer building, the behaviour of the ground appears less disturbed. The structure accommodates more easily the "large" deformation imposed by the ground. We can then deduce that it would be more difficult to reach the failure state of such a building.

In a similar manner, the behaviour of the PVC building model is easier to analyse thanks to the higher values of strains. Nevertheless within the different strain gauges, only half of them show a similar behaviour of the building between the two tests. Others do not allow to be exploited. The strains being larger, it tends to lesser the serviceability limit state of the building. More tests should be performed with this flexible model of building in order to determine more clearly its behaviour and also the influence of its stiffness in various positions relative to the cavity.

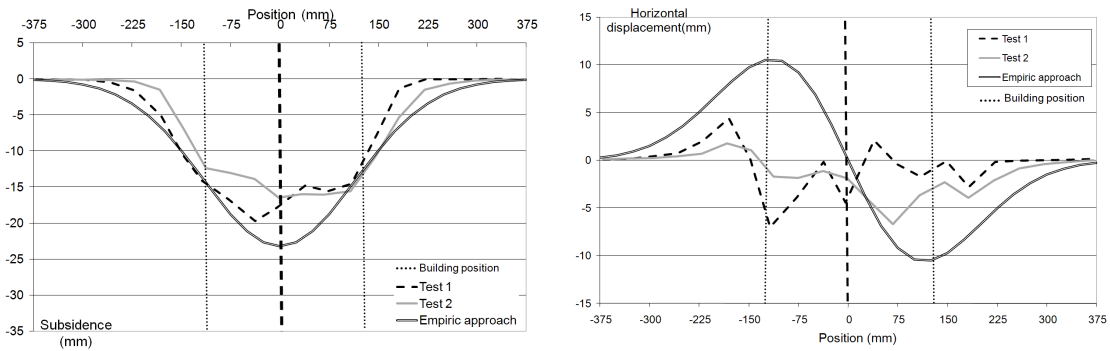

a) Ground movement for the position n°1 of the building (compression area)

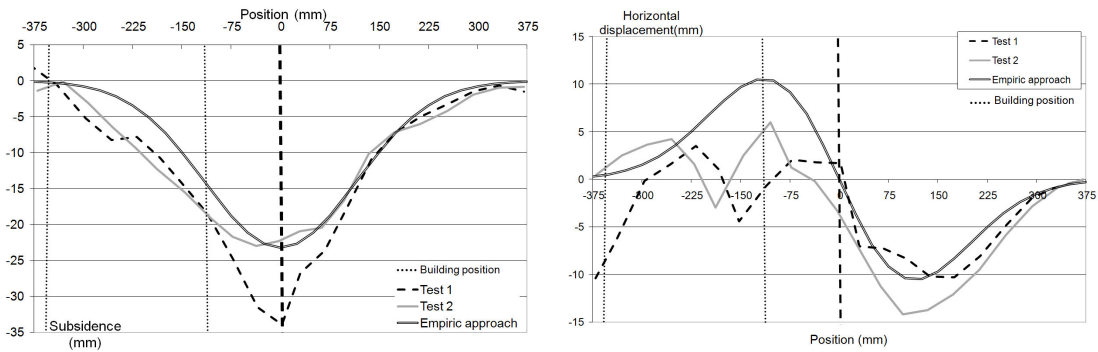

b) Ground movement for the position n°2 of the building (extension area)

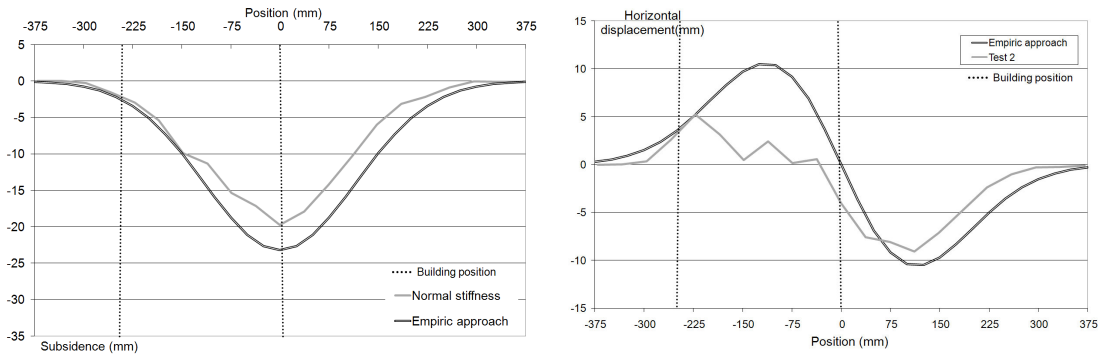

c) Ground movement for the position n°3 of the building (greatest slope area)

Figure 8. Comparison of the ground movements for the different positions of the building.

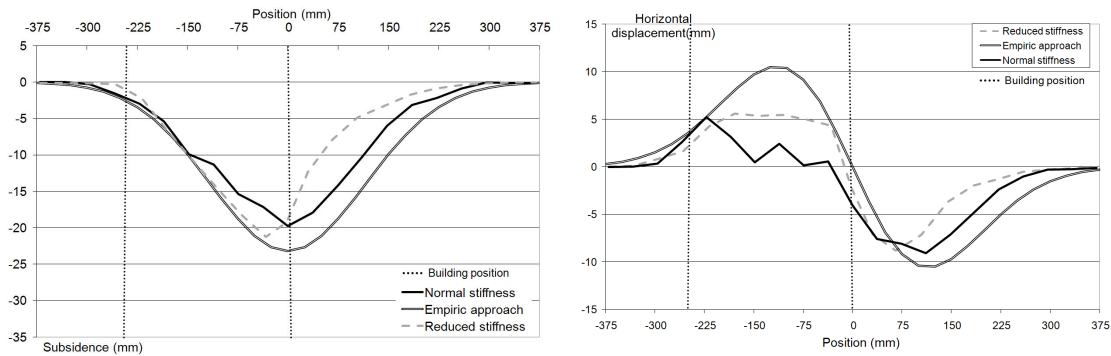

Figure 9. Comparison between the ground displacements for the two different stiffnesses of structure.

## 5 Conclusions and prospects

In order to synthesize the different test results, the table 3 presents them classified by position, stiffness of the building and then by nature of movements or deformations applied by the ground to the structure. The columns named "Greenfield" correspond to the hypothetic behaviour of the building subjected to the ground movement without taking into account the soil-structure interaction. Three normalized parameters (figure 10) are used to compare the different behaviours:

- $\varepsilon_h$: corresponds to the horizontal deformation on the whole structure;
- $\omega$: corresponds to the average slope applied by the ground to the structure;
- $\Delta$: is correlated to the curvature applied by the ground to the building model.

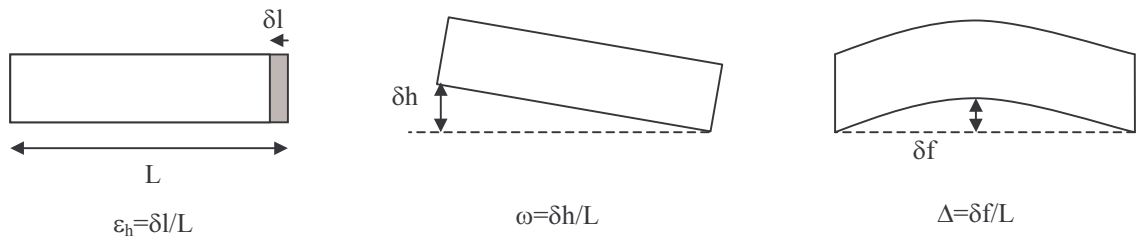

Figure 10. Explanations of the three parameters used to synthesize the behaviour of the building model.

Table 3. Summary of the results.

|  | Position n°1 | | Position n°2 | | Position n°3 | | |
| --- | --- | --- | --- | --- | --- | --- | --- |
|  | Greenfield | Building | Greenfield | Building | Greenfield | Building | Building reduced-stiffness |
| $\varepsilon_h$ | 4,17% | 0,31% | -7,97% | -0,69% | -0,77% | 0,08% | -1,50% |
| $\omega$ | 6,45% | 9,50% | 0,00% | 0,10% | 10,2% | 11,56% | 9,44% |
| $\Delta$ | 2,50% | 1,77% | 5,17% | 0,21% | 1,22% | 0,79% | 0,36% |

The effect of the stiffness of the building and its position is clearly seen on the horizontal deformations applied by the ground to the structure: a stiffer building is less concerned by the horizontal deformation. Regarding the slope, the movement of the building is always more important than the movement corresponding to the greenfield conditions (for position n°2, due to the uncertainties, the value measured with the building may be considered similar to the greenfield one). This is also true for the curvature of the building, it is less important than what would be expected if the soil-structure interaction was neglected.

However, the results concerning the tests using the building with a reduced stiffness are more difficult to understand, particularly the horizontal deformation. It does not seem logical that this building may present a horizontal deformation greater than the one corresponding to greenfield condition. This is observed on both tests using this setting. Future tests should be achieved in order to confirm this result and to determine the reasons of such behaviour.

The next step in the study of the influence of the soil-structure interaction on the consequences of ground movements on the structures would be to extend the parametric study. Other parameters should have a certain influence, such as the nature of the structure, its geometry and the depth of the cavity for example.

Two other ways of longer term research are considered: the evaluation of the very small movements of the building before the collapse of the cavity in order to predict its occurrence and the development of a three dimensional physical model.

While the LVDT sensors are used in this paper to measure the large displacements of the foundations of the building, it should be possible to use them to study the small movement happening during the few seconds preceding the collapse. Some information might be usable for some real cases and then allow to alert the population before a dangerous event.

INERIS is achieving the development of a three dimensional small-scale physical model allowing the study of structures and infrastructures submitted to various ground movements, from a local sinkhole to a subsidence of great extent. This tool should also allow in a close future to test different technical solutions for the protection of the buildings (peripheral trench, geosynthetics, ...).